\documentclass[twocolumn]{jetpl}

\usepackage{graphicx}
\usepackage{cite}
\usepackage{amsmath}
\usepackage{amsfonts}
\usepackage{color}

\begin{document}

\title{Critical behavior of transport and magnetotransport in 2D electron system
in Si in the vicinity of the metal-insulator transition}

\rtitle{Critical behavior of transport\ldots}

\sodtitle{Critical behavior of transport and magnetotransport in 2D electron system
in Si in the vicinity of metal-insulator transition}

\author{D.~A.~Knyazev, O.~E.~Omel'yanovskii, V.~M.~Pudalov, I.~S.~Burmistrov$^\dag\ddag$}

\rauthor{D.~A.~Knyazev, O.~E.~Omel'yanovskii, V.~M.~Pudalov, I.~S.~Burmistrov}

\sodauthor{D.~A.~Knyazev, O.~E.~Omelyanovskii, V.~M.~Pudalov, I.~S.~Burmistrov}

\address{P.~N.~Lebedev Physical Institute RAS, 119991 Moscow, Russia\\
$^\dag$ L.~D.~Landau Institute for Theoretical Physics RAS, 117940 Moscow, Russia\\
$\ddag$ Moscow Institute of Physics and Technology, Department of theoretical
physics, 141700 Moscow, Russia}
\dates{November 21, 2006}{*}

\abstract{We report on studies of the magnetoresistance in strongly correlated 2D electron system in Si
in the critical regime, in the close vicinity of the 2D metal-insulator transition.
We performed self-consistent comparison of our data with solutions of
two equation of the cross-over renormalization group (CRG) theory,
which describes temperature evolutions of the resistivity and
interaction parameters for 2D electron system. We found a good
agreement between the $\rho(T,B_\parallel)$ data and the RG theory
in a wide range of the in-plane fields, 0-2.1\,T. This agreement
supports the interpretation of the observed 2D MIT as the true
quantum phase transition.}


\PACS{73.40, 73.43}

\maketitle

Low-temperature transport in many high mobility 2D
electron systems was found to manifest a critical
behavior~\cite{prb94,prb95,review,varenna}. This phenomenon takes
place at low carrier densities, therefore the electron-electron correlations
play a crucial role. In transport studies, as temperature
decreases, the overall picture of the temperature dependence of
resistivity $\rho$ shows clearly distinct behavior in two
different domains on the density scale. At low densities, $n <
n_c$, resistivity exponentially increases with cooling, whereas at
high densities, $n > n_c$, resistivity significantly decreases
(here $n_c$ is a sample-dependent critical density, $\sim
10^{11}$\,cm$^{-2}$ for high mobility Si samples). In the former
domain, transport is consistent with a conventional picture of
hopping conduction, that is typical for an insulator~\cite{hop}.
In the latter domain, far from the critical density (and at $\rho
\ll h/e^2$), temperature dependence of conductivity was
experimentally shown \cite{quantcorr} to be explained by the
Fermi-liquid effects that were calculated within the framework
both of the quantum interaction corrections~\cite{zala}, and
temperature dependent screening~\cite{dsh}. It was so far unclear
whether the metallic type transport persists to $T = 0$ and
whether the cross-over from the metallic to the insulating
behavior (that is observed at low
though finite temperatures) signifies a true quantum phase
transition.

The successful comparison with the theory of interaction
corrections in the high density regime, $n\gg n_c$, encouraged us
to extend the comparison to the critical regime of lower densities
$n\approx n_c$ and higher resistivities $\rho \sim h/e^2$. The
method that now commonly in use for this regime and in the
diffusive interaction limit $T\tau \ll 1$ is a generalization of
the nonlinear $\sigma$-model theory, which has been developed by
Finkel'stein \cite{fink}. The RG equations
\cite{fink,castellani84} describe length scale (temperature)
evolutions of the resistivity and interaction parameters
for 2D electron system in the first order
in $\rho/(\pi h/e^2)$ and in all orders in interaction.

Earlier \cite{fink02},  only experimental $\rho(T)$-data
\cite{pudalov98} in zero field has been compared with one of the
RG-equations, while temperature dependence of the interaction
parameter $\gamma_2(T)$ \cite{fink} was not tested. In this paper
we have extracted, for the first time, $\gamma_2(T)$
 from the low-field magnetoresistance measurements, using cross-over RG (CRG) equations,
 proposed in Ref.~\cite{burm} for
in-plane magnetic fields, varying from vanishingly low to large.
We have compared $\gamma_2(T)$ with theoretical dependence
calculated from the RG-theory \cite{fink02} in zero field. Also,
we have compared our $\rho(T,B_\parallel)$ data with the solutions
of the CRG equations. For these purposes, we measured
$\rho(T,B_\parallel)$ in a wide range of the in-plane fields
 $B_\parallel=0-2.5$\,T.
We have found a good agreement of the measured
$\rho(T,B_\parallel)$ and $\gamma_2(T)$ with the RG theory.

Measurements were performed with a Si-MOS sample (peak mobility
3\,m$^2$/Vs at $T = 1.3$\,K) of the rectangular geometry
5$\times$0.8\,mm$^2$. We used four-terminal ac-technique at 5\,Hz
frequency; source-drain current  was chosen low enough,
$I=10$\,nA, in order to avoid electron overheating. The sample was
located at a precise rotation stage that enabled to align its
plane parallel to the magnetic field directions with accuracy of
$\sim 1'$. The alignment was controlled by observation of the
suppression of the weak localization (WL) peak in $\rho_{xx}$ in
the magnetic field $B$. For studies we have chosen the temperature
range $(1.3-4.2)$\,K, because for the studied high-mobility Si-MOS
sample in the critical regime $n\approx n_c$ the resistivity
exhibits a well-pronounced maximum at about $T_{\rm max}\approx 2
- 3$\,K  with relatively low resistivity $\rho_{\rm max} \sim
h/e^2$. For lower densities, the $\rho(T)$ maximum shifts to lower
temperatures and higher resistivities; as a results, $\rho_{\rm
max} (e^2/\pi h)$ is no longer a small parameter as needed for
comparison with the one-loop RG-theory.

\begin{figure}[h]
\centerline{\includegraphics[width=0.55\textwidth,height=0.55\textwidth]{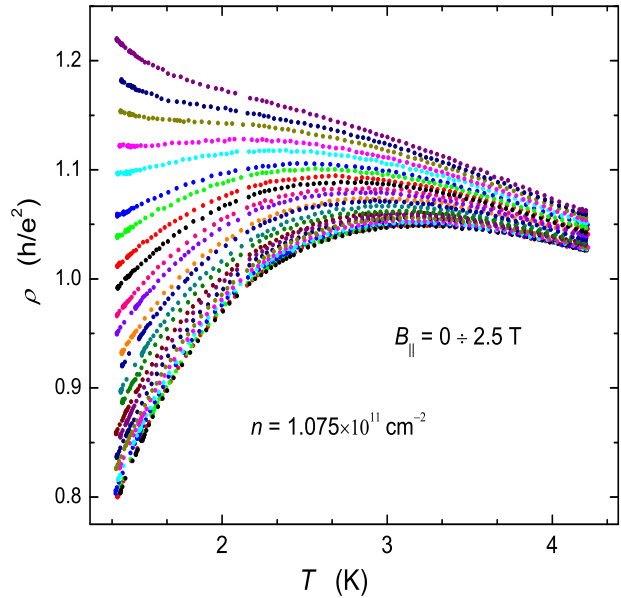}}
\caption{Fig.~1. $\rho(T)$ temperature dependences for various
in-plane magnetic fields varied  in steps
of 0.1\,T in the range $(0-2.5)$\,T (from bottom to top)}
\label{fig.1}
\end{figure}

Figure~\ref{fig.1} shows typical temperature dependences of the
resistivity in the critical regime for various $B_\parallel$
fields in the range from 0 to 2.5\,T. For the studied sample, the
metal-insulator transition in zero field takes place at the
critical density $n_c = 0.86\times10^{11}$\,cm$^{-2}$. Data shown
in Fig.~\ref{fig.1} were taken
 for the carrier density $n =
1.075\times 10^{11}$\,cm$^{-2}$ that is very close to $n_c$. There are several evidences
for the presented data belong to the critical regime: (i) the proximity of $n$ to $n_c$,
(ii) the non-monotonic $\rho(T)$ temperature dependence with a clearly
pronounced  maximum and with a  sharp drop in
$\rho(T)$ at temperatures below the maximum, and
(iii) the high  value of the low-temperature resistivity $\rho \approx h/e^2$,
that is close to the critical value
$\sim 2h/e^2$ for this sample.

Rough estimate of
the transport time $\tau$ based on the Drude
formula shows that
$T\tau$ is $\leq 0.3$ for all ``metallic-like'' curves (i.e. for
the curves with $\rho(T)$ maximum in Fig.~\ref{fig.1}), over the
whole studied temperature range. This indicates that the data shown in Fig.~1
belong to the diffusive interaction regime.

In zero magnetic fields, there is a well-pronounced maximum  in
$\rho(T)$ at  $T_{\rm max} \approx 3$\,K. Application of the
in-plane magnetic field gradually drives the system to the
insulating state. As $B_\parallel$ field increases, the  $\rho(T)$
maximum becomes shallow and shifts progressively to lower
temperatures (see Fig.~\ref{fig.1}).  At $B_\parallel= 2.3$\,T the
maximum completely vanishes; for higher fields resistivity
monotonically increases with cooling that signals the onset of the
insulating state. The cross-over from non-monotonic to monotonic
temperature dependence  of the  magnetoresistance occurs in fields
$g\mu_B B_\parallel \sim kT_{\rm max}$ in agreement with the
predictions of the CRG theory \cite{burm}.

\begin{figure}[h]
\centerline{\includegraphics[width=0.55\textwidth,height=0.53\textwidth]{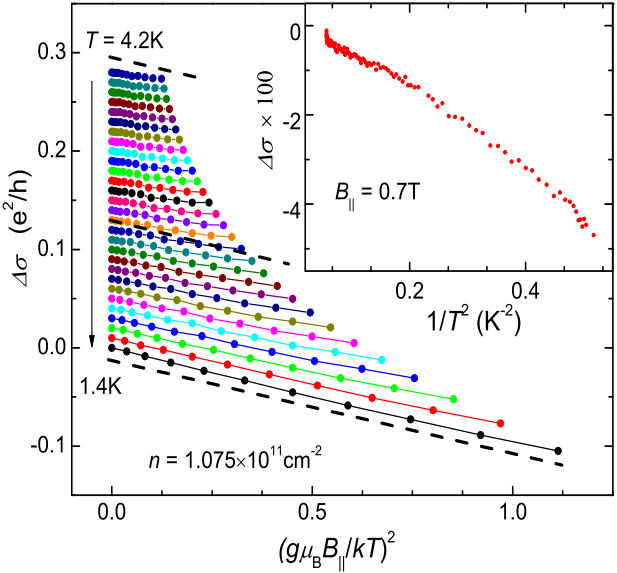}}
\caption{Fig.~2. Magnetoconductivity in the critical regime versus
square of the ratio of the Zeeman to thermal energy. Different
curves correspond to different temperatures,  varying,  from top
to bottom, in the range 4.2 to 1.4\,K. Dashed parallel lines are
guides to the eye. Inset shows magnetoconductivity at $B_\parallel
=0.7$\,T versus $1/T^2$}
\label{fig.2}
\end{figure}

Figure~\ref{fig.2} shows the magnetoconductivity data
$\Delta\sigma=\sigma(B_\parallel)-\sigma(0)$, replotted from
Fig.~\ref{fig.1} for various fixed temperatures and for
$B_\parallel$
varying from 0 to 1.1\,T. For low  fields,  $b=g\mu_B B_\parallel/kT
\ll 1$,  the data, as expected, are linear in $B_\parallel^2$. When
$\Delta\sigma(B)$ is plotted as a function of $b^2$,  the
slopes of the linear $|\Delta\sigma(B)|$ curves {\em increase as
temperature decreases} (compare the data with the guiding parallel dashed
lines in Fig.~\ref{fig.2}). This proves that the data are taken in
the diffusive interaction regime \cite{zala}. Indeed,
in the ballistic interaction regime, where the magnetoconductance
data scale as $(B^2/T)$ \cite{zala}, the slope would behave in the
opposite way, i.e. decrease linearly with cooling. The diffusive
interaction behavior can also be seen from the inset to
Fig.~\ref{fig.2}, where the same data for a fixed
$B_\parallel=0.7$\,T are plotted versus $(1/T)^2$. A positive
curvature of the data demonstrates that the $T$-dependence is even
steeper than $1/T^2$.

For the diffusive interaction regime, the first order interaction
corrections (which are applicable for low resistivity/high density
regime) result in the magnetoconductance as follows
\cite{CdCL,altsh}:
\begin{equation}
\sigma(T,B)=\sigma(T,0) - c n_v^2 \gamma_2(\gamma_2 + 1)\frac{(g
\mu_B B_\parallel)^2}{(kT)^2}, \label{CdCL}
\end{equation}
where $c \approx 0.029$, $\gamma_2 =-F_0^\sigma/(1+F_0^\sigma)$,
$F_0^\sigma$ - Fermi-liquid interaction constant, and $n_v^2$
takes into account correct number of triplet terms for the
multivalley system \cite{fink02}. Thus, within the framework of
the interaction corrections, the slope of the magnetoconductance
curves depends on the interaction parameter.

According to Eq.~(1), the decrease in slopes of the $\Delta\sigma(b)$ curves with
temperature in Fig.~\ref{fig.2} indicates a
temperature dependence of the interaction parameter, which is one
of the goals
of our studies. However,
Eq.~(\ref{CdCL}) is valid only for $\rho \ll h/e^2$, i.e. for high
density ``metallic'' regime, where temperature dependence of the
interaction constant should be logarithmically weak~\cite{altsh}.
On the other hand, our main interest is the critical regime of low
densities and high resistivities $\rho\sim h/e^2$.

\begin{figure}
\centerline{\includegraphics[width=0.55\textwidth,height=0.53\textwidth]{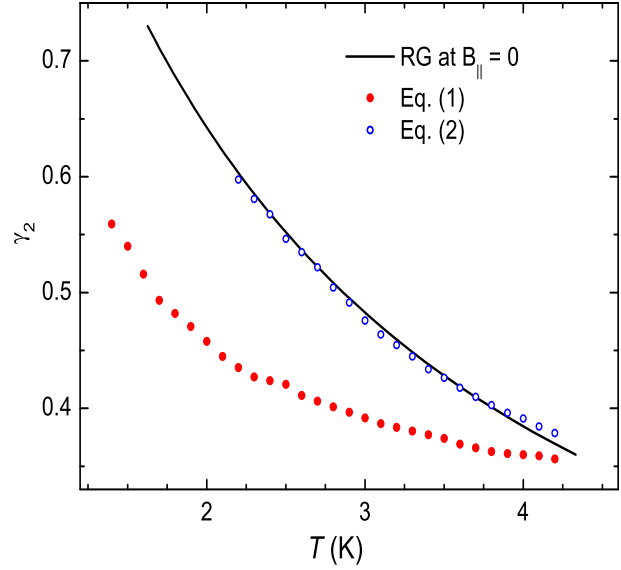}}
\vspace{-0.1in} \caption{Fig.~3. Temperature dependence of
$\gamma_2$. Solid line - theoretical curve calculated from the
RG-equations \protect\cite{fink02}. The symbols denote the
empirical $\gamma_2$ obtained from our  data using
Eq.~(\protect\ref{CdCL}) (filled circles), and using
Eq.~(\protect\ref{burm}) (empty circles)}
\label{fig.3}

\end{figure}

The two-parameter RG-theory has been developed to describe the
behavior of the 2D system in the critical regime
 for the zero
magnetic field case~\cite{fink,fink02}. It predicts a strong
(power-low) temperature dependence of $\gamma_2(T)$. This theory
was already shown \cite{fink02} to describe qualitatively the
nonmonotonic $\rho(T)$ experimental data in the critical regime.
In RG theory, the $\rho(T)$ maximum signifies a turnover from the
``localizing'' to ``delocalizing'' behavior. The $(\rho_{\rm
max},T_{\rm max})$ coordinates of this point, therefore, are
convenient for selecting the proper boundary conditions
(integration constants).
We have calculated numerically the $\gamma_2(T)$ dependence by solving
the system of RG-equations~\cite{fink02} in the one-loop
approximation for the two-valley case \cite{fink02}. This zero field result
is plotted in Fig.~\ref{fig.3} as a solid curve.

The experimental observation of such strong $\gamma_2(T)$
dependence would be a crucial test of the RG-theory. In order to
extract the $\gamma_2(T)$ dependence from magnetoconductance data,
a parallel magnetic field is evidently required. If magnetic field
is weak in all experimentally studied temperature range ($g\mu_B
B_\parallel \ll kT$) the dependence of the normalized resistance
$\rho(T,B_\parallel)/\rho(B_\parallel)_{\rm max}$ on $n_v
\rho(B_\parallel)_{\rm max} \ln T/T(B_\parallel)_{\rm max}$ is
given by the same universal curve as that for
$B_\parallel=0$~\cite{fink,burm}. Here $T(B_\parallel)_{\rm max}$
and $\rho(B_\parallel)_{\rm max}$ denote the temperature and the
resistivity maximum values at a given $B_\parallel$ field.
Expanding $T(B_\parallel)_{\rm max}$ and $\rho(B_\parallel)_{\rm
max}$ to the second order in $B_\parallel$ one can obtain a novel
expression for the magnetoresistivity:
\begin{equation}
\frac{\rho(T,B_\parallel)}{\rho(B_\parallel)_{\rm max}} =
\frac{\rho(T,0)}{\rho_{\rm max}} + \upsilon(\gamma_2) \frac{(g
\mu_B B_\parallel)^2}{(kT)^2}. \label{burm}
\end{equation}
Here the function $\upsilon(\gamma_2)$ depends not only on
$\gamma_2$ but on $T(B_\parallel)_{\rm max}$,
$\rho(B_\parallel)_{\rm max}$, as well as derivatives $d
T(B_\parallel)_{\rm max}/d b^2$ and $d \rho(B_\parallel)_{\rm
max}/d b^2$ at $b=0$.
In notations of Ref.~\protect\cite{burm}
\begin{eqnarray}
&\upsilon(\gamma_2)=a_0(\gamma_2)
e^{2F(\gamma_2)-2F(\gamma_2^\textrm{max})}\Bigl
[-\frac{d\ln\rho_\textrm{max}}{d b^2} \times\notag \\
&\int_{\gamma_2^\textrm{max}}^{\gamma_2} \frac{d u}{b_0(u)}
e^{-F(u)+F(\gamma_2^\textrm{max})} + \frac{\rho_\textrm{max}}{2}
\frac{d\ln T_\textrm{max}}{d b^2} \Bigr ]\Biggr |_{b=0}
\end{eqnarray}

To compare our data  in the low field limit $b\ll 1$ with Eq. (2),
for each temperature  we have determined  the difference
$\delta\rho=\rho(T,B_\parallel)/\rho(B_\parallel)_{\rm max}-
\rho(T,0)/\rho_{\rm max}$ and plotted it as function of $b^2$. The
$\upsilon(\gamma_2)$ function was calculated numerically, by
solving the CRG equations (17) and (18) from Ref.~\cite{burm} in
the   $b^2$ approximation. The slope of the resulting
$\delta\rho(b^2)$ curve enables us to extract the experimental
temperature dependence of $\gamma_2$ by using Eq.~(2). The result
is shown in Fig.~(3) by empty circles. One can see that our data
agrees with the theoretical $\gamma_2(T)$ dependence~\cite{fink02}
with no adjustable parameters in the wide range of temperatures.
We show in Fig.~3 the data only down to $T=2.2$\,K at which the
universal curve given by the one-loop RG theory starts to
overestimate $\rho(T,0)/\rho_{\rm max}$.
For lower temperatures, it is no longer possible to extract
$\gamma_2(T)$ from $\upsilon(\gamma_2)$ since the latter becomes
non-monotonic.
Presumably, it indicates the significance of higher-order (in
$\rho$) terms in the RG theory at such low temperatures.

For comparison, we also present in Fig.~(3) the $\gamma_2(T)$  dependence (filled circles)
determined from our experimental data by using Eq.~(1), i.e. the first-order interaction corrections.
There is only a qualitative similarity between thus determined $\gamma_2(T)$  and the
theoretical RG-result (solid line). The disagreement is not surprising because the
first-order (in $\rho$) interaction quantum corrections are inapplicable for the $\rho \sim h/e^2$ case.

It should be noted that the observed strong growth in
$\gamma_2(T)$ (see Fig.~3) at first sight is inconsistent with
almost temperature independent $F_0^\sigma$, determined from
Shubnikov-de Haas (ShdH) measurements \cite{SdH,granada}. There
are several possible reasons for this apparent inconsistency: (i)
a finite perpendicular and parallel fields, such as needed for the
observation of beats in ShdH measurements, affects strongly the
solution of the RG equations by reducing significantly
$\gamma_2(T)$, (ii) ShdH measurements of $F_0^\sigma$ might be
made away of the critical regime of electron densities, and (iii)
ShdH measurements at low densities were made at low temperatures
$T<1$\,K, whereas the critical $\gamma_2(T)$ behavior in Fig.~3 is
established reliably only for higher temperatures $T>2.2$\,K.

\begin{figure}
\centerline{\includegraphics[width=0.55\textwidth,height=0.53\textwidth]{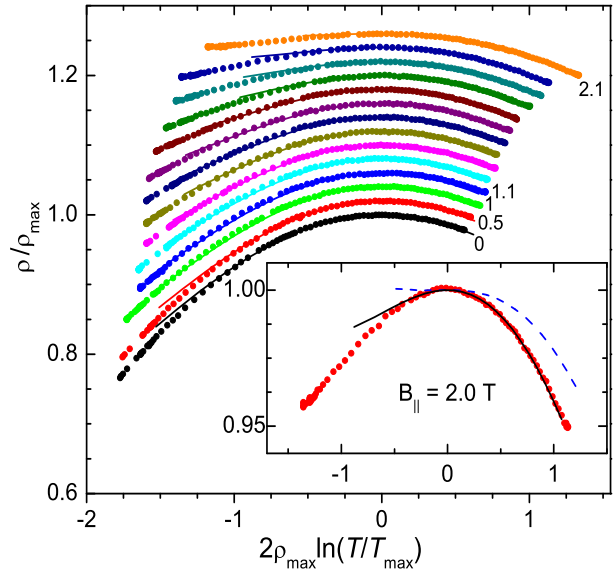}}
\caption{Fig.~4. Comparison of the measured
$\rho(T)$ data (symbols) with the theoretical dependences (lines)
calculated from RG-equations~\protect\cite{burm} for various fixed
magnetic fields (indicated next to each curve, in Tesla). Data and
curves are offset shifted vertically by 0.02, relative to each
other, starting from the zero field curve. Inset shows example of
the comparison between the $\rho(T)$ data  for $B_\parallel=2$\,T
and the CRG equations \protect\cite{burm} with $A=1/2\pi$ (dashed
curve) and $A=0.71/\pi$ (solid curve)}
\label{fig.4}
\end{figure}

In Ref.~\cite{burm} cross-over RG equations have been suggested
to describe the transition from weak ($b\ll
1$) to strong ($b\gg 1$) parallel magnetic fields. In order to
accomplish the comprehensive comparison of our data with the
theory, we have directly compared the measured
$\rho(T,B_\parallel)$ dependences with solutions of the cross-over
RG equations \cite{burm} for various fixed magnetic fields. For
this purpose, we normalized the $\rho(T)$ data by its maximum
value $\rho(B)_{\rm max}$ for each magnetic field. The comparison
is presented in Fig.~4. One can see the $\rho(T)$ data agree with
the RG theory not only in zero field (as was demonstrated earlier
\cite{fink02}), but also in the wide range of $B_\parallel$
fields.

In low fields $B_\parallel < 1.5$\,T there is a quantitative
agreement between the experiment and the RG theory. In fields $ <
0.7$\,T, as discussed above, the $\rho(T)$ data collapse onto the
universal curve for $B_\parallel=0$ with no adjustable parameters.
For higher magnetic field, the data collapse deteriorates because
the system enters the non-universal cross-over regime. In theory
\cite{burm}, magnetic field enters the cross-over RG equations as
\begin{equation}
\mathcal{B} = A [1+\gamma_2(T_\textrm{max})]\frac{g\mu_B
B_\parallel}{kT_{\rm max}}
\end{equation}
where the numerical prefactor $A=1/2\pi$.

In the cross-over regime for $B_\parallel >1.5$\,T, there is only
qualitative agreement between the theory \cite{burm} and the data
as shown in the inset to Fig.~4. However, by changing the
numerical factor $A$, we could bring the theoretical curves into a
good agreement with the data. For example, for $B_\parallel=2$\,T
the calculated curve agrees with the data when $A$ equals
$0.71/\pi$ (see Fig.~4). In even higher fields $B>2.3$\,T, the
$\rho(T)$ maximum vanishes and comparison with the theory is no
longer possible. We mention that in high fields the comparison
between the data and the RG theory is limited to not too low
temperatures due to the divergence of $\rho(T)$ (in theory) at
finite temperature in the one-loop CRG equations \cite{burm}.

Finally, we wish to  note that earlier \cite{fink02} only one of the RG-equations,
for $\rho(T)$ in zero field,
 has been tested by comparing with experimental data \cite{pudalov98}. In this comparison
  the temperature dependence of $\gamma_2(T)$
was taken solely from theory and other RG equations were not tested therefore.
Recently \cite{krav}, an attempt was undertaken to test both variables,
$\rho(T)$ and $\gamma_2(T)$, where the zero field $\rho(T)$-data  was compared with solution of the  RG-equation,
and $\gamma_2$ was obtained from the magnetoconductance data using Eq.~(1).
As discussed above,  Eq.~(1) describes interaction quantum corrections and, strictly speaking, is
inapplicable in the critical regime  of $\rho\sim h/e^2$. Indeed, Fig.~3 demonstrates that
the $\gamma_2(T)$ dependence, obtained in this way,  is only qualitatively similar to the
 exact solution of the RG-equations \cite{fink02}.

In contrast, in this paper,
we compared   self-consistently
both variables $\rho(T)$
and $\gamma_2(T)$ with solutions of two RG-equations \cite{fink02}
and found a good consistency  between the data and the theory.
Moreover, we have found a good agreement
of the $\rho(T,B_\parallel)$ data with the solutions of the
CRG-equations \cite{burm} for various $B_\parallel$ fields.

In conclusion, we performed a comprehensive comparison of the
transport and magnetotransport critical behavior in a wide
temperature range, with the cross-over RG-equations which take
magnetic field into account. Specifically, we compared with the
theory temperature dependences of both, the interaction constant
$\gamma_2$ and resistivity $\rho$. We have found that (i) the
experimental $\gamma_2(T)$ determined from the low-field
magnetoresistance  grows rapidly as temperature decreases, in
agreement  with calculated $\gamma_2(T)$~\cite{fink02}, and (ii)
the calculated temperature dependence of the magnetoresistance is
in a good agreement with the experimental $\rho(T,B_\parallel)$
data in a wide magnetic field range. This agreement strongly
supports the theoretical interpretation of the observed 2D MIT as
the true quantum phase transition.

The authors are grateful  to A.~M.~Finkelstein for
discussions, and V.~S.~Tur for the excellent technical assistance.
The work was partially supported by RFBR, Programs of RAS and the
Russian Ministry for education and science. D.A.K. and I.S.B. acknowledge
grants from the Russian Science Support Foundation, and
the Council For Grants of the President of Russian Federation.


\begin{thebibliography}{99}

\bibitem{prb94}
S.~V.~Kravchenko, G.~V.~Kravchenko, J.~E.~Furneaux, V.~M.~Pudalov, and M.~D'Iorio,
Phys. Rev. B {\bf 50} 8039 (1994).

\bibitem{prb95}
S.~V.~Kravchenko, W.~E.~Mason, G.~E.~Bowker, J.~E.~Furneaux,
V.~M.~Pudalov, and M.~D'Iorio, Phys. Rev. B {\bf 51} 7038 (1995).

\bibitem{review}
for the review see,
E.~Abrahams, S.~V.~Kravchenko, M.~P.~Sarachik, Rev. Mod. Phys. {\bf 73} 251 (2001).

\bibitem{varenna}
for the review see, V.~M.~Pudalov,
in: {\em The Electron Liquid Paradigm in Condensed Matter Physics}, Ed.
by G.~F.~Giuliani, and G.~Vignale (IOS press, Amsterdam, 2004),
p.335-356. cond-mat/0405315.

\bibitem{hop}
B.~I.~Shklovskii, A.~L.~Efros, Electronic Properties of Doped Semiconductors in
Springer Series in Solid State Sciences 45 (Springer Verlag, New York) 1984.


\bibitem{quantcorr}
Y.~Y.~Proskuryakov, A.~K.~Savchenko, S.~S.~Safonov,
M. Pepper, M.Y. Simons, and D.A. Ritchie,
Phys. Rev. Lett. {\bf 89} 076406 (2002). A.~A.~Shashkin,
S.~V.~Kravchenko, V.~T.~Dolgopolov, and T.~M.~Klapwijk,
Phys. Rev. B {\bf 66} 076303 (2002). S.~A.~Vitkalov, K.~James,
B.~N.~Narozhny, M.P. Sarachik, and T.M. Klapwijk,
Phys. Rev. B {\bf 67} 113310 (2003). V.~M.~Pudalov,
M.~E.~Gershenson, H.~Kojima, G.~Brunthaler, A. ~Prinz, and G.~Bauer,
Phys. Rev. Lett. {\bf 91} 126403 (2003).


\bibitem{zala}
G.~Zala, B.~N.~Narozhny, and I.~L.~Aleiner, Phys. Rev. B {\bf 64} 214204 (2001);
Phys. Rev. B {\bf 65} 020201 (2001).

\bibitem{dsh}
S.~Das~Sarma and E.~H.~Hwang, Phys. Rev. B {\bf 69} 195305 (2004);
Phys. Rev. Lett. {\bf 83} 164 (1999).

\bibitem{fink}
A.~M.~Finkelstein, Z. Phys. B {\bf 56} 189 (1984); {\em Electron liquid in disordered
conductors}, Sov. Phys.
Reviews, Sec. A, ed. by I.~M.~Khalatnikov (Harwood Academic Publishers, London, 1990)  {\bf 14}, p.1.

\bibitem{castellani84}
C.~Castellani, C.~Di Castro, P.~A.~Lee, and M.~Ma, Phys. Rev. B
{\bf 30} 527 (1984).

\bibitem{fink02}
A.~Punnoose and A.~M.~Finkel'stein, Phys. Rev. Lett. {\bf 88}
16802 (2002).

\bibitem{pudalov98}
V. M. Pudalov, G. Brunthaler, A. Prinz, and G. Bauer, Pis'ma v
ZhETF {\bf 68} 415 (1998);
[JETP Lett. {\bf 68}, 442 (1998)].

\bibitem{burm}Preceding Letter:
I.~S.~Burmistrov, N.~M.~Chtchelkatchev, Pis'ma v ZhETF {\bf 84},
775 (2006).

\bibitem{CdCL}
C.~Castellani, C.~Di~Castro, P.~A.~Lee, Phys. Rev. B {\bf 57} 9381
(1998).

\bibitem{altsh}
B.~L.~Al'tshuler, A.~G.~Aronov, in: {\em Electron-electron
interactions in disordered systems} ed. by A.~L.~Efros and
M.~Pollack (North-Holland, Amsterdam, 1985), p. 1-154.

\bibitem{SdH}V.\,M.\,Pudalov, M.\,E.\,Gershenson, H.\,Kojima, N.\,Butch,
E.\,M.\,Dizhur, G.\,Brunthaler, A.\,Prinz, G.\,Bauer,
Phys. Rev. Lett. {\bf 88}, 196404 (2002).

\bibitem{granada} V.~M.~Pudalov, M.~E.~Gershenson, H.~Kojima,
Chapter 19 in: {\em Fundamental problems of mesoscopic physics},
NATO science series v.{\bf 154}, Ed. by I. Lerner, B. Altshuler, and
Y.Gefen, (Kluwer Academic Publishers, Dordrecht, 2004), p.309-327.
cond-mat/0401396.

\bibitem{krav}
S.~Anissimova, S.~V.~Kravchenko, A.~Punnoose, A.~M.~Finkel'stein, and
T.~M.~Klapwijk, cond-mat/0609181.



\end{thebibliography}
\end{document}